\documentclass[%
print,
 amsmath,amssymb,
 aps
]{revtex4}

\usepackage{times}%此指令是令文章字体选用Palation.
\usepackage{graphicx}% Include figure files
\usepackage{subfigure}
\usepackage{dcolumn}% Align table columns on decimal point
\usepackage{bm}% bold math
\usepackage{dsfont}
\usepackage{mathrsfs}
\usepackage[landscape,papersize={297.1mm,210mm},left=1.9cm,right=1.6cm,top=2.7cm,bottom=2.8cm]{geometry}% 此指令是规定A4纸的页边距
\usepackage[                   %dvipdfm, pdflatex,pdftex这里决定运行文件的方式不同
            pdfstartview=FitH,
            colorlinks, %注释掉此项则交叉引用为彩色边框(将colorlinks和pdfborder同时注释掉)
            pdfborder=001,   %注释掉此项则交叉引用为彩色边框
            linkcolor=blue,
            anchorcolor=blue,
            citecolor=blue,
            urlcolor=blue
            ]{hyperref}
\usepackage{amsthm}
\usepackage{pifont}
\usepackage{diagbox}
\usepackage{multirow}
\makeatletter

\newcommand{\Rmnum}[1]{\expandafter\@slowromancap\romannumeral #1@}

\makeatother

\begin{document}
\title{ Quantifying nonclassical correlation via the generalized Wigner-Yanase skew information}

\author{Yan Hong$^{1,2}$}

\author{Xinlan Hao$^{1}$}

\author{Limin Gao$^{1,2}$}
\email{gaoliminabc@163.com (corresponding author)}
\affiliation{$^1$ School of Mathematics and Science,  Hebei GEO University, Shijiazhuang 050031,  China \\
 $^2$   Intelligent Sensor Network Engineering Research Center of Hebei Province, Hebei GEO University, Shijiazhuang 050031,  China}

\begin{abstract}
Nonclassical correlation is an important concept in quantum information theory, referring to a special type of correlation that exists between quantum systems, which surpasses the scope of classical physics. In this paper, we introduce the concept of a family of information with important properties, namely the generalized Wigner-Yanase skew information, of which the famous quantum Fisher information and  Wigner-Yanase-Dyson skew information are special cases. We classify the local observables into two categories (i.e., orthonormal bases and Hermitian operators with a fixed nondegenerate spectrum), and based on this, we propose several indicators to quantify nonclassical correlation of bipartite quantum states. We have not only investigated some important properties of these  indicators but also illustrated through specific examples that they can indeed capture nonclassical correlation. Furthermore, we find that these indicators reduce to  entanglement measure for bipartite pure states. Specifically, we also derive the relationship between these indicators and the entanglement measure known as $I$-concurrence.
\end{abstract}

\maketitle

\section{Introduction}

The correlation is an important physical resource with significant applications in quantum information processing and quantum computing, and the characterization and quantification of correlation is an important research topic in quantum information theory \cite{HorodeckiRMP2009,GuhneToth2009,ModiBrodutchCablePaterekVedral2012,ReidDrummondBowenCavalcantiLamBachorAndersenLeuchs2009,UolaCostaNguyenGuhe2020}.
So far, there have been various methods to witness all kinds of correlations, such as detecting entanglement through quantum Fisher information and skew information \cite{Chen71052302,Hyllus2012,Goth2012,LiLuo2013PRA,HongLuoSong2015,ZhangFei2019,AkbariAzhdargalam2019,AnnaBoschiPRB2023,HongXingGaoGaoYan2024}, and detecting steering by  uncertainty relations \cite{WalbornSallesGomesToscanoRibeiro2011,SchneelochBroadbentWalbornCavalcantiHowell2013,JiLeeParkNha2015,CostaUolaGuhne2018,LaiLuo2022,GiananiBerardiBarbieri2022,LiFanSongYeWang2024}.

The depiction of correlation from different points of view has led to various definitions of correlation, such as entanglement \cite{HorodeckiRMP2009,BennettDiVincenzoSmolinWootters1996,Vidal2000}, quantum discord \cite{OllivierZurek2001,Luo2008,YuWuWangYiSong2014},  measurement-induced nonlocality \cite{LuoFu2011,HuFan2015,MuthuganesanSankaranarayanan2017}.
As one of the most renowned correlations, entanglement has been extensively studied, and there are various measures to quantify it, such as  entanglement cost \cite{PlenioVirmani2007}, entanglement distillation \cite{PlenioVirmani2007}, entanglement of formation \cite{WoottersPRL1998}, concurrence \cite{WoottersPRL1998}, and so on. Quantum steering and Bell nonlocality
are quantum correlations that are stronger than entanglement, and they have attracted widespread attention and research \cite{UolaCostaNguyenGuhe2020,WisemanJonesDoherty2007,Bell1964,RevModPhys81865,ReidDrummondBowenCavalcantiLamBachorAndersenLeuchs2009,FanZhaoMingWangYe2023}.

Based on the Wigner-Yanase skew information, Ref. \cite{LuoFu2012} introduced a measure for correlation which was defined with the difference between the information content of  $\rho_{12}$ and $\rho_1\otimes \rho_2$ for bipartite quantum system $\mathcal{H}_{1}\otimes \mathcal{H}_{2}$, and this measure can be straightforwardly calculated. And a great advantage of this measure is that it does not involve optimization problems and can be directly calculated.
Based on Wigner-Yanase-Dyson skew information, this  method was similarly promoted, and the promoted quality satisfied some properties of correlation measure \cite{FanWang2020}.
For bipartite state $\rho_{12}$, the relative quantum Fisher information  was introduced as a measure of correlation,  which was the difference of quantum Fisher information between $\rho_{12}$ and $\rho_1\otimes \rho_2$ \cite{XingChenFu2019}.
The relative quantum Fisher information not only satisfied the conditions of  a correlation measure, but also reduced to an entanglement measure for pure states \cite{XingChenFu2019}.

For bipartite quantum systems, Ref. \cite{KimLiKumarWu2018} defined two different ways of nonclassical correlation measures in terms of quantum Fisher information, and pointed out that these two correlation measure were consistent with the geometric discord for pure states.
However, the contractivity of the measure involving local measurements under local completely positive and trace-preserving  map has not been proven, and it remains an unresolved issue.
Girolami $et$ $al$.  proposed a nonclassical correlation measure based on Wigner-Yanase skew information, proved that this measure reduced to an entanglement measure for pure states, and gave the analytical expression of this measure for the qubit-qudit system \cite{GirolamiTufarelliAdesso2013}.
Gibilisco $et$ $al$.  generalized this measure  by  the metric adjusted skew information,
and the extended nonclassical correlation measure still satisfied some of the superior properties of the original measure \cite{GibiliscoGirolamiHansen2021}.

Quantum Fisher information and Wigner-Yanase skew information play  important roles in the quantification of correlation.
In this paper, we will characterize the nonclassical correlation through the generalized Wigner-Yanase skew information and the  metric-adjusted skew information.
In Sec. II,  we will introduce the generalized Wigner-Yanase skew information and the  metric-adjusted skew information, which include some important information such as quantum Fisher information and Wigner-Yanase skew information, and prove that these information content have some important properties.
In Sec. III, we will present some different  indicators for quantifying nonclassical correlation  and demonstrate that they possess some desirable properties.
In Sec. IV, we will specifically elaborate on the relationship between these  indicators and entanglement measure.
In Sec. V, we will  summarize this paper.

\section{the generalized Wigner-Yanase skew information and the  metric-adjusted skew information}

Let us introduce the following function with $a\geq0,b\geq0, s\leq0, 0<\omega<1$,
\begin{equation*}
\begin{aligned}
 f^\omega_s(a,b)=\left\{\begin{array}{ll} \big[\omega a^s+(1-\omega)b^s\big]^{1/s}, &  \textrm{ for } ab\neq0,\\\\
0, &  \textrm{ for } ab=0,
\end{array}
\right.
\end{aligned}
\end{equation*}
when $0<\omega<1$ and $s\in(-\infty,0)$;
\begin{equation*}
\begin{aligned}
f^\omega_0(a,b)=&\lim\limits_{s\rightarrow0}f^\omega_s(a,b)=\left\{\begin{array}{ll} a^\omega b^{1-\omega}, &  \textrm{ for } ab\neq0,\\\\
0, &  \textrm{ for } ab=0 ,
\end{array}
\right.
\end{aligned}
\end{equation*}
when $0<\omega<1$ and $s=0$;
\begin{equation*}
\begin{aligned}
f^\omega_{-\infty}(a,b)=&\lim\limits_{s\rightarrow-\infty}f^\omega_s(a,b)
=&\left\{\begin{array}{ll} \min\{a,b\}, &  \textrm{ for } ab\neq0,\\
0, &   \textrm{ for }   ab=0,
\end{array}
\right.
\end{aligned}
\end{equation*}
when $0<\omega<1$ and $s=-\infty$.

Let the spectral decomposition of quantum state $\rho$ be $\rho=\sum\limits_i\lambda_i|\psi_i\rangle\langle\psi_i|$,
the generalized Wigner-Yanase skew information of the observable $X$ in the state $\rho$ is defined as
\begin{equation*}
\begin{aligned}
I^\omega_s(\rho,X):=\textrm{Tr}(\rho X^2)-\sum\limits_{i, j}f^\omega_s(\lambda_i,\lambda_{j})|\langle\psi_i|X|\psi_{j}\rangle|^2.
\end{aligned}
\end{equation*}
With the following derivation,
\begin{equation*}
\begin{aligned}
\textrm{Tr}(\rho X^2)=\sum\limits_i\lambda_i\langle\psi_i| X^2|\psi_{i}\rangle
=\sum\limits_i\lambda_i\langle\psi_i|X(\sum\limits_j|\psi_{j}\rangle\langle\psi_j|)X|\psi_{i}\rangle
=\sum\limits_{i,j}\lambda_i|\langle\psi_i|X|\psi_{j}\rangle|^2,
\end{aligned}
\end{equation*}
the generalized Wigner-Yanase skew information can be expressed as
\begin{equation*}
\begin{aligned}
I^\omega_s(\rho,X)=\sum\limits_{i\neq j}[\lambda_i-f^\omega_s(\lambda_i,\lambda_{j})]|\langle\psi_i|X|\psi_{j}\rangle|^2.
\end{aligned}
\end{equation*}
When $s=0$, the generalized Wigner-Yanase skew information $I^\omega_0(\rho,X)$ is Wigner-Yanase-Dyson skew information \cite{Lieb1973}.
When $\omega=\frac{1}{2}$, the generalized Wigner-Yanase skew information $I^{\frac{1}{2}}_s(\rho,X)$ is the information introduced by Ref. \cite{YangQiao2022}.
Specifically, for $\omega=\frac{1}{2}$, when $s=0$ or $s=-1$, $I^{\frac{1}{2}}_0(\rho,X)$ and $I^{\frac{1}{2}}_{-1}(\rho,X)$ represent Wigner-Yanase skew information and quantum Fisher information, respectively.

The  metric-adjusted skew information  can be written as
\begin{equation*}
\begin{aligned}
F_f(\rho,X)=\frac{f(0)}{2}\sum\limits_{i, j}\dfrac{(\lambda_i-\lambda_{j})^2}{\lambda_jf(\lambda_i/\lambda_j)}|\langle\psi_i|X|\psi_{j}\rangle|^2,
\end{aligned}
\end{equation*}
where the operator monotone function ($f$ is
called operator monotone if $A\leq B$ implies that $f (A) \leq f (B)$
for any Hermitian operators  $A$ and $B$) $f: \textbf{R}^+\longrightarrow \textbf{R}^+$ satisfies $f(0)>0$
and $xf(1/x)=f (x)$. When $f(x)=\dfrac{\omega(1-\omega)(x-1)^2}{(x^\omega-1)(x^{1-\omega}-1)}$ or $f(x)=\dfrac{1+x}{2}$, the metric-adjusted skew information are reduced to Wigner-Yanase-Dyson skew information and quantum Fisher information \cite{FrankHansen2008,CaiHansen2010}, respectively.

The generalized Wigner-Yanase skew information and the  metric-adjusted skew information have some important properties  as follows:

(1) (Monotonicity of $s$) For any observable $X$, the generalized Wigner-Yanase skew information $I^\omega_{s}(\rho,X)$ is  monotonically decreasing with  $s$, that is,
\begin{equation*}
\begin{aligned}
 I^\omega_0(\rho,X)\leq\cdots\leq I^\omega_{-\infty}(\rho,X).
 \end{aligned}
\end{equation*}
The equality holds when $\rho$  is a pure state. When the quantum state is pure state, $I^\omega_{s}(|\psi\rangle\langle\psi|,X)= V(|\psi\rangle\langle\psi|,X)=F_f(|\psi\rangle\langle\psi|,X)$ for any $s$ and $0<\omega<1$.

(2) (Non-negativity) $I^\omega_{s}(\rho,X)\geq0$. $I^\omega_{s}(\rho,X)=0$ if and only if $\rho$ and $X$ are commutative.
 $F_f(\rho,X)\geq0$ \cite{CaiHansen2010}. $F_f(\rho,X)=0$ if and only if $\rho$ and $X$ are commutative \cite{CaiHansen2010}.

(3) (Invariant under unitary transformations) $I^\omega_{s}(\rho,X)$ is  unitary invariant,
$I^\omega_{s}(U\rho U^\dagger,X)=I^\omega_{s}(\rho,U^\dagger XU)$ for any unitary matrix $U$.
 $F_f(\rho,X)$  is  unitary invariant \cite{CaiHansen2010},
 $F_f(U\rho U^\dagger,X)=F_f(\rho,U^\dagger XU)$ for any unitary matrix $U$.

(4) (Convexity) For any observable $X$, the  metric-adjusted skew information is convex, that is,
\begin{equation*}
\begin{aligned}
F_f(\rho,X)\leq\sum\limits_ip_iF_f(\rho_i,X),
\end{aligned}
\end{equation*}
where $\rho=\sum\limits_ip_i\rho_i$.
 For any observable $X$, when $-1\leq s\leq0$, $I^\omega_{s}(\rho,X)$ is convex  in a certain sense, that is,
\begin{equation*}
\begin{aligned}
I^\omega_{s}(\rho,X)\leq\sum\limits_jq_jI^\omega_{s}(|\phi_j\rangle\langle\phi_j|,X)
\end{aligned}
\end{equation*}
when $-1\leq s\leq0$, where $\rho=\sum\limits_jq_j|\phi_j\rangle\langle\phi_j|$.

(5) (Additivity) For a bipartite quantum  state $\rho_1\otimes\rho_2\in\mathcal{H}_1\otimes \mathcal{H}_2$, the generalized Wigner-Yanase skew information holds
\begin{equation*}
\begin{aligned}
I^\omega_s(\rho_1\otimes\rho_2,X_1\otimes \textbf{1}_2+\textbf{1}_1\otimes X_2 )=I^\omega_s(\rho_1,X_1)+I^\omega_s(\rho_2,X_2),
\end{aligned}
\end{equation*}
the  metric-adjusted skew information holds \cite{CaiHansen2010}
\begin{equation*}
\begin{aligned}
F_f(\rho_1\otimes\rho_2,X_1\otimes \textbf{1}_2+\textbf{1}_1\otimes X_2 )=F_f(\rho_1,X_1)+F_f(\rho_2,X_2),
\end{aligned}
\end{equation*}
where $\rho_i$ is the quantum state on subsystem $\mathcal{H}_i$, $X_i$ is an observable acting on the subsystem $\mathcal{H}_i$ and $\textbf{1}_j$ is the identity matrix acting on the subsystem $\mathcal{H}_j$.

The proofs of  the properties for  the generalized Wigner-Yanase skew information are placed in Appendix A.

\section{ the characterization of non-classical correlations}

Next, we specify the local observables in the generalized Wigner-Yanase skew information and the metric-adjusted skew information to two classes of
measurements: orthonormal bases,  a Hermitian operator with fixed nondegenerate spectrum.

\subsection{The nonclassical correlation involving  orthonormal bases }

Since $I^\omega_{s}(\rho,X)\geq0$ and $F_f(\rho,X)\geq$ for any Hermite operator $X$,
we define  indicators for a bipartite state $\rho$ of the quantum system $\mathcal{H}_{1}\otimes \mathcal{H}_{2}$ with dim$\mathcal{H}_{i}=d_i$  as follows
\begin{equation}\label{correlation1}
\begin{aligned}
 \mathcal{I}^\omega_s(\rho):=\min\limits_{\{|\chi_l\rangle\}}\Big[\sum\limits_{l=1}^{d_1} I^\omega_s(\rho,|\chi_l\rangle\langle\chi_l|\otimes \textbf{1}_2)\Big]^{\frac{1}{2}},
\end{aligned}
\end{equation}
\begin{equation}\label{correlation11}
\begin{aligned}
\mathcal{F}_f(\rho):=\min\limits_{\{|\chi_l\rangle\}}\Big[\sum\limits_{l=1}^{d_1} F_f(\rho,|\chi_l\rangle\langle\chi_l|\otimes \textbf{1}_2)\Big]^{\frac{1}{2}},
\end{aligned}
\end{equation}
where the minimum    runs over all possible orthonormal bases $\{|\chi_l\rangle:l=1,2,\cdots,d_1\}$ of subsystems  $\mathcal{H}_{1}$.
For special cases with $s=-1$ and $\omega=\frac{1}{2}$,  the indicator $\min\limits_{\{|\chi_l\rangle\}}\sum\limits_{l=1}^{d_1} I^{\frac{1}{2}}_{-1}(\rho,|\chi_l\rangle\langle\chi_l|\otimes \textbf{1}_2)$ has already been introduced in Ref. \cite{KimLiKumarWu2018}, which corresponds to $\Big(\mathcal{I}^{\frac{1}{2}}_{-1}(\rho)\Big)^2$.

$\emph{Theorem 1.} $   The indicators $ \mathcal{I}^\omega_s(\rho)$  defined as  (\ref{correlation1}) and $\mathcal{F}_f(\rho)$ defined as  (\ref{correlation11}) have the following properties.

(P1) $\rho$ is a classical-quantum state if and only if  $\mathcal{I}^\omega_s(\rho)=0$ (or $\mathcal{F}_f(\rho)=0$).

 (P2) $\mathcal{I}^\omega_s(\rho)$ and $\mathcal{F}_f(\rho)$ are locally unitary invariant, that is,
\begin{equation*}
\begin{aligned}
\mathcal{I}^\omega_s(\rho)=\mathcal{I}^\omega_s\Big((U_1\otimes U_2)\rho (U_1^\dagger\otimes U_2^\dagger)\Big),
\mathcal{F}_f(\rho)=\mathcal{F}_f\Big((U_1\otimes U_2)\rho (U_1^\dagger\otimes U_2^\dagger)\Big),
\end{aligned}
\end{equation*}
where $U_1$ and $U_2$ are any unitary operators acting on subsystem $\mathcal{H}_{1}$  and  $\mathcal{H}_{2}$, respectively.

 (P3a) $\mathcal{F}_f(\rho)$ is decreasing under the local completely positive and trace-preserving (CPTP) map $I_1\otimes\varepsilon_2$,
\begin{equation*}
\begin{aligned}
 \mathcal{F}_f(I_1\otimes\varepsilon_2(\rho))\leq \mathcal{F}_f(\rho).
\end{aligned}
\end{equation*}
When $s=0$ or when $\omega=\frac{1}{2}$ and $s=-1$, $\mathcal{I}^\omega_s(\rho)$ is decreasing under the local completely positive and trace-preserving (CPTP) map $I_1\otimes\varepsilon_2$,
  \begin{equation*}
\begin{aligned}
\mathcal{I}^\omega_s(I_1\otimes\varepsilon_2(\rho))\leq \mathcal{I}^\omega_s(\rho).
\end{aligned}
\end{equation*}
Here $I_1$ being the identity operation on subsystem $\mathcal{H}_{1}$  and $\varepsilon_2$ being the CPTP map on  subsystem $\mathcal{H}_{2}$,

(P3b)  If $\rho_2=\textrm{tr}_1(\rho)$ is the maximally mixed state,  $\mathcal{I}^\omega_s(\rho)$ is decreasing under random unitary channel $\varepsilon_2$ on  subsystem $\mathcal{H}_{2}$  when $s=0$ or when $\omega=\frac{1}{2}$ and $ s=-1$ and $F_f(\rho)$ is also decreasing under random unitary channel $\varepsilon_2$.

\emph{Proof.} (P1)
If $\rho$ is a classical-quantum state, then it can be written $\rho=\sum\limits_{l=1}^m\lambda_l|\phi_l\rangle\langle\phi_l|\otimes\tau_l$
with  $\langle\phi_l|\phi_{l'}\rangle=\delta_{ll'}$. When $m=d_1$,  $\{|\phi_l\rangle: l=1, 2, \cdots, d_1\}$ is precisely an orthonormal basis of subsystem  $\mathcal{H}_{1}$. When $m<d_1$, we can find $d_1-m$ pure states, denoted as $|\phi_{m+1}\rangle,|\phi_{m+2}\rangle,\cdots,|\phi_{d_1}\rangle$, so that  $\{|\phi_l\rangle: l=1, 2, \cdots, d_1\}$ is an orthonormal basis of subsystem  $\mathcal{H}_{1}$.
We can easily verify that $|\phi_l\rangle\langle\phi_l|\otimes \textbf{1}_2$ and $\rho$ are commutative for any $l=1,2,\cdots,d_1$, so $\sum\limits_{l=1}^{d_1} I^\omega_s(\rho,|\phi_l\rangle\langle\phi_l|\otimes \textbf{1}_2)=0$ and $\sum\limits_{l=1}^{d_1} F_f(\rho,|\phi_l\rangle\langle\phi_l|\otimes \textbf{1}_2)=0$  under  orthonormal basis $\{|\phi_l\rangle: l=1, 2, \cdots, d_1\}$  of subsystem  $\mathcal{H}_{1}$. This indicates that $\mathcal{I}^\omega_s(\rho)=0$ and $\mathcal{F}_f(\rho)=0$.

 If $\mathcal{I}^\omega_s(\rho)=0$, there exits an orthonormal basis $\{|\chi_l\rangle:l=1,2,\cdots,d_1\}$ of subsystem  $\mathcal{H}_{1}$ so that
$\sum\limits_{l=1}^{d_1}I^\omega_s(\rho,|\chi_l\rangle\langle\chi_l|\otimes \textbf{1}_2)=0$, so $I^\omega_s(\rho,|\chi_l\rangle\langle\chi_l|\otimes \textbf{1}_2)=0$ for any $l=1, 2, \cdots, d_1$. Hence,  we can obtain that $\rho$ and $|\chi_l\rangle\langle\chi_l|\otimes \textbf{1}_2$ are commutative for any $l=1, 2, \cdots, d_1$.
And then  quantum state $\rho$ owns spectral decomposition $\rho=\sum\limits_{l=1}^{d_1}\sum\limits_{l'=1}^{d_2}\lambda_{ll'}|\chi_l\rangle\langle\chi_l|\otimes|\omega_{ll'}\rangle\langle\omega_{ll'}|$ with
$\{|\chi_l\rangle|\omega_{ll'}\rangle:l=1,2,\cdots,d_1,l'=1,2,\cdots,d_2\}$ being the set of eigenvectors of $|\chi_l\rangle\langle\chi_l|\otimes \textbf{1}_2$. We can see that quantum state $\rho$ is a classical-quantum state because it follows from
\begin{equation*}
\begin{aligned}
\rho=&\sum\limits_{l=1}^{d_1}\sum\limits_{l'=1}^{d_2}\lambda_{ll'}|\chi_l\rangle\langle\chi_l|\otimes|\omega_{ll'}\rangle\langle\omega_{ll'}|\\
=&\sum\limits_{l=1}^{d_1}(\sum\limits_{l'=1}^{d_2}\lambda_{ll'})|\chi_l\rangle\langle\chi_l|\otimes
(\sum\limits_{l'=1}^{d_2}\dfrac{\lambda_{ll'}}{\sum\limits_{l'=1}^{d_2}\lambda_{ll'}}|\omega_{ll'}\rangle\langle\omega_{ll'}|).
\end{aligned}
\end{equation*}
Similarly, we can prove that $\rho$ is a classical-quantum state if $\mathcal{F}_f(\rho)=0$.

(P2) Let us prove  that $\mathcal{I}^\omega_s(\rho)$  is  locally unitary invariant.
\begin{align}
& \mathcal{I}^\omega_s\big((U_1\otimes U_2)\rho (U_1^\dagger\otimes U_2^\dagger)\big)\label{}\nonumber\\
=&\min\limits_{\{|\chi_l\rangle\}}\Big[\sum\limits_{l=1}^{d_1} I^\omega_s\big((U_1\otimes U_2)\rho (U_1^\dagger\otimes U_2^\dagger),|\chi_l\rangle\langle\chi_l|\otimes \textbf{1}_2\big)\Big]^{\frac{1}{2}}\label{P1.1}\\
=&\min\limits_{\{|\chi_l\rangle\}}\Big[\sum\limits_{l=1}^{d_1} I^\omega_s\big(\rho, (U_1^\dagger|\chi_l\rangle\langle\chi_l|U_1)\otimes \textbf{1}_2)\big)\Big]^{\frac{1}{2}}\label{P1.2}\\
=&\mathcal{I}^\omega_s(\rho).\label{P1.3}
\end{align}
Here Eq. (\ref{P1.1}) holds by the definitions  (\ref{correlation1}). Eq. (\ref{P1.2})  is true because  $I^\omega_{s}(\rho,X)$  is unitary invariant. Eq. (\ref{P1.3}) is valid as the minimum is over all orthonormal bases of subsystems  $\mathcal{H}_{1}$ and any orthonormal basis is still an orthonormal basis after unitary transformation.
Similarly, we can prove that $\mathcal{F}_f(\rho)$  is  locally unitary invariant by (\ref{correlation11}) and the facts $F_f(\rho,X)$ is  unitary invariant and any orthonormal basis is still an orthonormal basis after unitary transformation.

(P3a) For a bipartite quantum state $\rho$, suppose that $\{|\widetilde{\chi}_l\rangle\}$ is the optimal orthonormal basis that achieves the minimum in the definition $\mathcal{F}_f(\rho)$, we can get
\begin{equation*}
\begin{aligned}
 \mathcal{F}_f(\rho)=&\Big[\sum\limits_{l=1}^{d_1} F_f(\rho,|\widetilde{\chi}_l\rangle\langle\widetilde{\chi}_l|\otimes \textbf{1}_2)\Big]^{\frac{1}{2}}\nonumber\\
\geq&\Big[\sum\limits_{l=1}^{d_1} F_f(I_1\otimes\varepsilon_2(\rho),|\widetilde{\chi}_l\rangle\langle\widetilde{\chi}_l|\otimes \textbf{1}_2)\Big]^{\frac{1}{2}}\nonumber\\
\geq&\min\limits_{\{|\chi_l\rangle\}}\Big[\sum\limits_{l=1}^{d_1} F_f(I_1\otimes\varepsilon_2(\rho),|\chi_l\rangle\langle\chi_l|\otimes \textbf{1}_2)\Big]^{\frac{1}{2}}\nonumber\\
=&\mathcal{F}_f(I_1\otimes\varepsilon_2(\rho)),
\end{aligned}
\end{equation*}
where the first inequality holds because  the metric-adjusted skew information is decreasing under  local CPTP map \cite{GibiliscoGirolamiHansen2021}, that is,
\begin{equation}\label{}\nonumber
\begin{aligned}
F_f(\rho,X_1\otimes \textbf{1}_2))\geq F_f(I_1\otimes\varepsilon_2(\rho),X_1\otimes \textbf{1}_2).
\end{aligned}
\end{equation}
When $f(x)=\dfrac{\omega(1-\omega)(x-1)^2}{(x^\omega-1)(x^{1-\omega}-1)}$ or $f(x)=\dfrac{1+x}{2}$, the metric-adjusted skew information are reduced to Wigner-Yanase-Dyson skew information and quantum Fisher information \cite{FrankHansen2008,CaiHansen2010}, respectively.
Hence,  $\mathcal{I}^\omega_s(\rho)$ is decreasing under  local CPTP map when $s=0$ or when $\omega=\frac{1}{2}$ and $s=-1$ because $\mathcal{F}_f(\rho)$ is decreasing under  local CPTP map.

(P3b) Since $\varepsilon_2$ is a random unitary operation on the state space of $\mathcal{H}_2$,
we have$$\varepsilon_2(\rho_2)=\sum\limits_{k} p_{k}U_{k}\rho_2U_{k}^{\dagger}$$ where $U_{k}$ is the  unitary operation on $\mathcal{H}_2$, and  $0\leq p_{k}\leq1, \sum\limits_{k} p_{k}=1$ \cite{GregorattiGregoratti2003,LandauStreater1993}.
 When $s=0$ or when $\omega=\frac{1}{2}$ and $ s=-1$,
 we have
\begin{align}
&\mathcal{I}^\omega_s(I_1\otimes\varepsilon_2(\rho))\nonumber\\
=&\min\limits_{\{|\chi_l\rangle\}}\Big[\sum\limits_{l=1}^{d_1}I^\omega_{s}(\sum\limits_{k} p_{k}(\textbf{1}_1\otimes U_{k})\rho(\textbf{1}_1\otimes U_{k})^{\dagger},|\chi_l\rangle\langle\chi_l|\otimes \textbf{1}_2)\Big]^{\frac{1}{2}}\nonumber\\
\leq&\min\limits_{\{|\chi_l\rangle\}}\Big[\sum\limits_{l=1}^{d_1}\sum\limits_{k} p_{k}I^\omega_{s}((\textbf{1}_1\otimes U_{k})\rho(\textbf{1}_1\otimes U_{k})^{\dagger},|\chi_l\rangle\langle\chi_l|\otimes \textbf{1}_2)\Big]^{\frac{1}{2}}\nonumber\\
=&\min\limits_{\{|\chi_l\rangle\}}\Big[\sum\limits_{l=1}^{d_1}\sum\limits_{k} p_{k}I^\omega_{s}(\rho,|\chi_l\rangle\langle\chi_l|\otimes \textbf{1}_2)\Big]^{\frac{1}{2}}\nonumber\\
=&\min\limits_{\{|\chi_l\rangle\}}\Big[\sum\limits_{l=1}^{d_1}I^\omega_{s}(\rho,|\chi_l\rangle\langle\chi_l|\otimes \textbf{1}_2)\Big]^{\frac{1}{2}}\nonumber\\
=&\mathcal{I}^\omega_s(\rho),\nonumber
\end{align}
 by the convexity of Wigner-Yanase-Dyson skew information ($s=0$) and quantum Fisher information ($\omega=\frac{1}{2}$ and $ s=-1$),  and  the fact that the generalized Wigner-Yanase skew information is a unitary invariant.
Similarly, we can also derive that $\mathcal{F}_f(I_1\otimes\varepsilon_2(\rho))\leq\mathcal{F}_f(\rho)$ by the convexity of the metric-adjusted skew information  and  the fact that the metric-adjusted skew information is a unitary invariant.

\subsection{The nonclassical correlation  involving a Hermitian operator with fixed nondegenerate spectrum}
In Ref. \cite{GirolamiTufarelliAdesso2013}, an  nonclassical correlation measure based on skew information has been presented
\begin{equation}\label{}\nonumber
\begin{aligned}
\mathfrak{I}(\rho):=\min\limits_{\{\Pi_\chi\}} I(\rho,\Pi_\chi\otimes \textbf{1}_2),
\end{aligned}
\end{equation}
and in Ref. \cite{GibiliscoGirolamiHansen2021}, an  nonclassical correlation measure based on metric-adjusted skew information has been given as follows
\begin{equation}\label{}\nonumber
\begin{aligned}
\mathfrak{F}_f(\rho):=\min\limits_{\{\Pi_\chi\}} F_f(\rho,\Pi_\chi\otimes \textbf{1}_2).
\end{aligned}
\end{equation}
Here the minimum runs over all possible Hermitian operators $\Pi_\chi$ with fixed nondegenerate spectrum  $\chi$ in subsystem  $\mathcal{H}_{1}$.
The Ref. \cite{GirolamiTufarelliAdesso2013} (or \cite{GibiliscoGirolamiHansen2021}) has been proven  that $\mathfrak{I}(\rho)$ (or $\mathfrak{F}_f(\rho)$) satisfies:
 If $\rho$ is a classical-quantum state if and only if  $\mathfrak{I}(\rho)=0$ (or $\mathfrak{F}_f(\rho)=0$); $\mathfrak{I}(\rho)$ (or $\mathfrak{F}_f(\rho)$) is locally unitary invariant;
$\mathfrak{I}(\rho)$ (or $\mathfrak{F}_f(\rho)$) is decreasing under the local completely positive and trace-preserving (CPTP) map.

For a bipartite state $\rho$ of the quantum system $\mathcal{H}_{1}\otimes \mathcal{H}_{2}$ with dim$\mathcal{H}_{i}=d_i$, we define another indicator in terms of the generalized Wigner-Yanase skew information as follows
\begin{equation}\label{correlation2}
\begin{aligned}
\mathcal{I}^\chi_{\omega,s}(\rho):=\min\limits_{\{\Pi_\chi\}} I^\omega_s(\rho,\Pi_\chi\otimes \textbf{1}_2),
\end{aligned}
\end{equation}
where the minimum runs over all possible Hermitian operators $\Pi_\chi$ with fixed nondegenerate spectrum  $\chi$ in subsystem  $\mathcal{H}_{1}$.
For special cases with $s=0$ and $\omega=\frac{1}{2}$,  the indicator $\mathcal{I}^{\frac{1}{2}}_{0}(\rho)$ is precisely  $\mathfrak{I}(\rho)$.

$\emph{Theorem 2.}$   The indicator $ \mathcal{I}^\chi_{\omega,s}(\rho)$  defined as  (\ref{correlation2}) is a correlation measure has the following properties.

 (PI)  $\rho$ is a classical-quantum state if and only if $\mathcal{I}^\chi_{\omega,s}(\rho)=0$.

 (PII) $\mathcal{I}^\chi_{\omega,s}(\rho)$ is locally unitary invariant.

 (PIIIa) When $s=0$ or when $\omega=\frac{1}{2}$ and $s=-1$, $\mathcal{I}^\chi_{\omega,s}(\rho)$ is decreasing under  local CPTP map.
(PIIIb) When $s=0$ or when $\omega=\frac{1}{2}$ and $s=-1$, if $\rho_2=\textrm{tr}_1(\rho)$ is the maximally mixed state,  $\mathcal{I}^\chi_{\omega,s}$ is decreasing under random unitary channel $\varepsilon_2$ on  subsystem $\mathcal{H}_{2}$.

\emph{Proof.} (PI)
If $\rho$ is a classical-quantum state, then it can be written $\rho=\sum\limits_{l=1}^m\lambda_l|\phi_l\rangle\langle\phi_l|\otimes\tau_l$
with  $\langle\phi_l|\phi_{l'}\rangle=\delta_{ll'}$. When $m=d_1$,  $\{|\phi_l\rangle: l=1, 2, \cdots, d_1\}$ is precisely an orthonormal basis of subsystem  $\mathcal{H}_{1}$. When $m<d_1$, we can find $d_1-m$ pure states, denoted as $|\phi_{m+1}\rangle,|\phi_{m+2}\rangle,\cdots,|\phi_{d_1}\rangle$, so that  $\{|\phi_l\rangle: l=1, 2, \cdots, d_1\}$ is an orthonormal basis of subsystem  $\mathcal{H}_{1}$.
We can easily verify that $\sum\limits_{l=1}^{d_1}\chi_l|\phi_l\rangle\langle\phi_l|\otimes \textbf{1}_2$ and $\rho$ are commutative, so $ I^\omega_s(\rho, \Pi_\chi\otimes \textbf{1}_2)=0$ with $\Pi_\chi=\sum\limits_{l=1}^{d_1}\chi_l|\phi_l\rangle\langle\phi_l|$. This indicates that $\mathcal{I}^\chi_{\omega,s}(\rho)=0$.

If $\mathcal{I}^\chi_{\omega,s}(\rho)=0$, there exits a Hermitian operators $\widetilde{\Pi}_\chi$ with fixed nondegenerate spectrum  $\chi=\{\chi_{1},\chi_{2},\cdots,\chi_{d_1}\}$ in subsystem  $\mathcal{H}_{1}$ so that
$\mathcal{I}^\chi_{\omega,s}(\rho)=I^\omega_s(\rho,\widetilde{\Pi}_\chi\otimes \textbf{1}_2)=0$.
Hence, we can obtain that $\rho$ and $\widetilde{\Pi}_\chi\otimes \textbf{1}_2$ are commutative.
Therefore,  quantum state $\rho$ owns spectral decomposition $\rho=\sum\limits_{l=1}^{d_1}\sum\limits_{l'=1}^{d_2}\lambda_{ll'}|\varphi_l\rangle\langle\varphi_l|\otimes|\omega_{ll'}\rangle\langle\omega_{ll'}|$ with
$\{|\varphi_l\rangle:l=1,2,\cdots,d_1\}$ being an unique  eigenbasis (up to the global phase factor) of $\widetilde{\Pi}_\chi$.
Using the derivation method of (P1) in Theorem 1, $\rho=\sum\limits_{l=1}^{d_1}\sum\limits_{l'=1}^{d_2}\lambda_{ll'}|\varphi_l\rangle\langle\varphi_l|\otimes|\omega_{ll'}\rangle\langle\omega_{ll'}|$  is a classical-quantum state.

(PII) The indicator $\mathcal{I}^\chi_{\omega,s}(\rho)$ remains unchanged under local unitary transformations, because
\begin{align}
& \mathcal{I}^\chi_{\omega,s}\Big((U_1\otimes U_2)\rho (U_1^\dagger\otimes U_2^\dagger)\Big)\label{}\nonumber\\
=&\min\limits_{\{\Pi_\chi\}} I^\omega_s\Big((U_1\otimes U_2)\rho (U_1^\dagger\otimes U_2^\dagger),\Pi_\chi\otimes \textbf{1}_2\Big)\label{}\nonumber\\
=&\min\limits_{\{\Pi_\chi\}} I^\omega_s\Big(\rho, (U_1^\dagger\Pi_\chi U_1)\otimes \textbf{1}_2)\Big)\label{}\nonumber\\
=&\mathcal{I}^\chi_{\omega,s}(\rho).\label{P2.1}
\end{align}
Here  Eq. (\ref{P2.1}) is valid as the minimum is over all Hermitian operators $\Pi_\chi$ with fixed spectrum  $\chi$ and the eigenvalues of an operator remain invariant under a unitary transformation.

(PIIIa) When $f(x)=\dfrac{\omega(1-\omega)(x-1)^2}{(x^\omega-1)(x^{1-\omega}-1)}$ or $f(x)=\dfrac{1+x}{2}$, the metric-adjusted skew information are reduced to Wigner-Yanase-Dyson skew information and quantum Fisher information \cite{FrankHansen2008,CaiHansen2010}, respectively.
Hence, the item (PIIIa) holds  because the metric-adjusted skew information is decreasing under the local completely positive and trace-preserving (CPTP) map \cite{GibiliscoGirolamiHansen2021}.

(PIIIb) Using a derivation approach similar to that of the property (P3b) in Theorem 1, we can also derive  as follows,
\begin{align}
&\mathcal{I}^\chi_{\omega,s}(I_1\otimes\varepsilon_2(\rho))\nonumber\\
=&\min\limits_{\{\Pi_\chi\}}I^\omega_{s}(\sum\limits_{k} p_{k}(\textbf{1}_1\otimes U_{k})\rho(\textbf{1}_1\otimes U_{k})^{\dagger},\Pi_\chi\otimes \textbf{1}_2)\nonumber\\
\leq&\min\limits_{\{\Pi_\chi\}}\sum\limits_{k} p_{k}I^\omega_{s}((\textbf{1}_1\otimes U_{k})\rho(\textbf{1}_1\otimes U_{k})^{\dagger},\Pi_\chi\otimes \textbf{1}_2)\nonumber\\
=&\min\limits_{\{\Pi_\chi\}}\sum\limits_{k} p_{k}I^\omega_{s}(\rho,\Pi_\chi\otimes \textbf{1}_2)\nonumber\\
=&\min\limits_{\{\Pi_\chi\}}I^\omega_{s}(\rho,\Pi_\chi\otimes \textbf{1}_2)\nonumber\\
=&\mathcal{I}^\chi_{\omega,s}(\rho),\nonumber
\end{align}
when $s=0$ or when $\omega=\frac{1}{2}$ and $ s=-1$.

\section{ Examples }

\emph{Example 1.} For a bipartite pure state $\rho=|\psi\rangle\langle\psi|$ of the quantum system $\mathcal{H}_{1}\otimes \mathcal{H}_{2}$ with dim$\mathcal{H}_{i}=d_i$, let the Schmidt decomposition of pure state $|\psi\rangle$ be $|\psi\rangle=\sum\limits_i\lambda_i|i_1\rangle|i_2\rangle$.

From Ref. \cite{KimLiKumarWu2018}, for the pure state $|\psi\rangle=\sum\limits_i\lambda_i|i_1\rangle|i_2\rangle$,
 $\Big(\mathcal{I}^{\frac{1}{2}}_{-1}(|\psi\rangle\langle\psi|)\Big)^2=\min\limits_{\{|\chi_l\rangle\}}\sum\limits_{l=1}^{d_1} I^{\frac{1}{2}}_{-1}(|\psi\rangle\langle\psi|,|\chi_l\rangle\langle\chi_l|\otimes \textbf{1}_2)=1-\sum\limits_i\lambda_i^4$ where the minimum    runs over all possible orthonormal bases $\{|\chi_l\rangle:l=1,2,\cdots,d_1\}$ of subsystems  $\mathcal{H}_{1}$.
 So, we have $\min\limits_{\{|\chi_l\rangle\}}\sum\limits_{l=1}^{d_1} I^{\omega}_{s}(|\psi\rangle\langle\psi|,|\chi_l\rangle\langle\chi_l|\otimes \textbf{1}_2)=\min\limits_{\{|\chi_l\rangle\}}\sum\limits_{l=1}^{d_1} F_f(|\psi\rangle\langle\psi|,|\chi_l\rangle\langle\chi_l|\otimes \textbf{1}_2)=1-\sum\limits_i\lambda_i^4$
because of   $I^\omega_{s}(|\psi\rangle\langle\psi|,X)=F_f(|\psi\rangle\langle\psi|,X)$ are equal for any $s\leq0,0<\omega<1$. Hence,  $\mathcal{I}^\omega_s(|\psi\rangle\langle\psi|)=\mathcal{F}_f(|\psi\rangle\langle\psi|)=\sqrt{1-\sum\limits_i\lambda_i^4}$.

Especially, let us consider the bipartite pure state $\rho=|\psi\rangle\langle\psi|$ of the quantum system $\mathcal{H}_{1}\otimes \mathcal{H}_{2}$ with dim$\mathcal{H}_{1}=2$. From Ref. \cite{GirolamiTufarelliAdesso2013}, when the nondegenerate spectrum  $\chi$ of subsystems  $\mathcal{H}_{1}$ is $\chi=\{-1,1\}$, $\mathcal{I}^\chi_{{\frac{1}{2}},0}(|\psi\rangle\langle\psi|)=\min\limits_{\{\Pi_\chi\}} I^{\frac{1}{2}}_0(|\psi\rangle\langle\psi|,\Pi_\chi\otimes \textbf{1}_2)=2(1-\textrm{Tr}\rho_1^2)$ where the minimum runs over all possible Hermitian operators $\Pi_\chi$ with fixed nondegenerate spectrum  $\chi$ in subsystems  $\mathcal{H}_{1}$. Therefore, we can also get $\mathcal{I}^\chi_{s,\omega}(|\psi\rangle\langle\psi|)=2(1-\textrm{Tr}\rho_1^2)=2(1-\sum\limits_{i=1}^2\lambda_i^4)$ with  $\chi=\{-1,1\}$ and dim$\mathcal{H}_{1}=2$.

\emph{Example 2.} For a bipartite quantum state
\begin{equation}\label{example20}
\begin{aligned}
\rho=\lambda|\psi_1\rangle\langle\psi_1|+(1-\lambda)|\psi_2\rangle\langle\psi_2|
 \end{aligned}
\end{equation}
 of the quantum system $\mathcal{H}_{1}\otimes \mathcal{H}_{2}$ with dim$\mathcal{H}_{1}=$  dim$\mathcal{H}_{2}=2$.
Here $|\psi_1\rangle=\frac{|00\rangle+|11\rangle}{\sqrt{2}}$ and $|\psi_2\rangle=\frac{|00\rangle-|11\rangle}{\sqrt{2}}$.

For the indicator $\mathcal{I}^\omega_s(\rho)$ defined as (\ref{correlation1}),  we can get
\begin{equation}\label{example21}
\begin{aligned}
 \mathcal{I}^\omega_s(\rho)=&\min\limits_{\{|\chi_l\rangle\}}\Big[1-\lambda\sum\limits_{l=1}^{2}|\langle\psi_1|(|\chi_l\rangle\langle\chi_l|\otimes \textbf{1}_2)|\psi_1\rangle|^2
-(1-\lambda)\sum\limits_{l=1}^{2}|\langle\psi_2|(|\chi_l\rangle\langle\chi_l|\otimes \textbf{1}_2)|\psi_2\rangle|^2\\
-&f^\omega_s(\lambda,1-\lambda)\sum\limits_{l=1}^{2}|\langle\psi_1|(|\chi_l\rangle\langle\chi_l|\otimes \textbf{1}_2)|\psi_2\rangle|^2
-f^\omega_s(1-\lambda,\lambda)\sum\limits_{l=1}^{2}|\langle\psi_2|(|\chi_l\rangle\langle\chi_l|\otimes \textbf{1}_2)|\psi_1\rangle|^2 \Big]^{\frac{1}{2}}.
\end{aligned}
\end{equation}
For the indicator $\mathcal{F}_f(\rho)$ defined as (\ref{correlation11}),  we can get
\begin{equation}\label{example211}
\begin{aligned}
 \mathcal{F}^2_f(\rho)=&f(0)\min\limits_{\{|\chi_l\rangle\}}\sum\limits_{l=1}^{2}\Big(\dfrac{(2\lambda-1)^2}{\lambda f(\frac{1-\lambda}{\lambda})}|\langle\psi_1|(|\chi_l\rangle\langle\chi_l|\otimes \textbf{1}_2)|\psi_2\rangle|^2+\dfrac{\lambda^2}{0f(\frac{\lambda}{0})}|\langle\psi_1|(|\chi_l\rangle\langle\chi_l|\otimes \textbf{1}_2)|\psi_3\rangle|^2\\
 =&+\dfrac{\lambda^2}{0f(\frac{\lambda}{0})}|\langle\psi_1|(|\chi_l\rangle\langle\chi_l|\otimes \textbf{1}_2)|\psi_4\rangle|^2
 +\dfrac{(1-\lambda)^2}{0f(\frac{1-\lambda}{0})}|\langle\psi_2|(|\chi_l\rangle\langle\chi_l|\otimes \textbf{1}_2)|\psi_3\rangle|^2
 +\dfrac{(1-\lambda)^2}{0f(\frac{1-\lambda}{0})}|\langle\psi_2|(|\chi_l\rangle\langle\chi_l|\otimes \textbf{1}_2)|\psi_4\rangle|^2
 \Big),
\end{aligned}
\end{equation}
where $|\psi_3\rangle=\frac{|01\rangle+|10\rangle}{\sqrt{2}}$ and $|\psi_4\rangle=\frac{|01\rangle-|10\rangle}{\sqrt{2}}$.

After calculation, we have
\begin{equation}\label{example22}
\begin{aligned}
\langle\psi_1|(|\chi_l\rangle\langle\chi_l|\otimes\textbf{1}_2)|\psi_1\rangle
=\langle\psi_2|(|\chi_l\rangle\langle\chi_l|\otimes \textbf{1}_2)|\psi_2\rangle=\dfrac{1}{2},
\end{aligned}
\end{equation}
\begin{equation}\label{example23}
\begin{aligned}
\langle\psi_1|(|\chi_l\rangle\langle\chi_l|\otimes\textbf{1}_2)|\psi_2\rangle
=\dfrac{|\langle\chi_l|0\rangle|^2-|\langle\chi_l|1\rangle|^2}{2}
=\dfrac{2a_l-1}{2},
\end{aligned}
\end{equation}
\begin{equation}\label{example231}
\begin{aligned}
|\langle\psi_1|(|\chi_l\rangle\langle\chi_l|\otimes\textbf{1}_2)|\psi_3\rangle|^2=|\langle\psi_2|(|\chi_l\rangle\langle\chi_l|\otimes\textbf{1}_2)|\psi_4\rangle|^2
=\dfrac{|\langle\chi_l|1\rangle\langle0|\chi_l\rangle+\langle\chi_l|0\rangle\langle1|\chi_l\rangle|^2}{4},
\end{aligned}
\end{equation}
\begin{equation}\label{example232}
\begin{aligned}
|\langle\psi_1|(|\chi_l\rangle\langle\chi_l|\otimes\textbf{1}_2)|\psi_4\rangle|^2
=|\langle\psi_2|(|\chi_l\rangle\langle\chi_l|\otimes\textbf{1}_2)|\psi_3\rangle|^2
=\dfrac{|\langle\chi_l|1\rangle\langle0|\chi_l\rangle-\langle\chi_l|0\rangle\langle1|\chi_l\rangle|^2}{4},
\end{aligned}
\end{equation}
with $a_l=|\langle\chi_l|0\rangle|^2$.
By substituting Eqs. (\ref{example22}) and (\ref{example23}) into Eq. (\ref{example21}), we can obtain
\begin{equation}\label{example24}
\begin{aligned}
 \mathcal{I}^\omega_s(\rho)=\Big[\frac{1}{2}-\frac{f^\omega_s(\lambda,1-\lambda)+f^\omega_s(1-\lambda,\lambda)}{4}
\max\limits_{\{|\chi_l\rangle\}} \sum\limits_{l=1}^{2}(2a_l-1)^2 \Big]^{\frac{1}{2}}.
\end{aligned}
\end{equation}
Using $\sum\limits_{l=1}^{2}a_l=\sum\limits_{l=1}^{2}\langle 0|\chi_l\rangle\langle \chi_l|0\rangle=1$, one has
\begin{equation}\label{example25}
\begin{aligned}
\max\limits_{\{|\chi_l\rangle\}} \sum\limits_{l=1}^{2}(2a_l-1)^2=2,
\end{aligned}
\end{equation}
where the maximum is  achieved when $a_1=0,a_2=1$ or $a_1=1,a_2=0$, that is, the maximum is  achieved under orthonormal basis $\{|0\rangle,|1\rangle\}$ of subsystem  $\mathcal{H}_{1}$.
By utilizing Eqs. (\ref{example24}) and (\ref{example25}), we are able to derive $\mathcal{I}^\omega_s(\rho)$ of $\rho$ defined as (\ref{example20}),
\begin{equation*}
\begin{aligned}
 \mathcal{I}^\omega_s(\rho)=&\min\limits_{\{|\chi_l\rangle\}}\Big[\sum\limits_{l=1}^{2} I^\omega_s(\rho,|\chi_l\rangle\langle\chi_l|\otimes \textbf{1}_2)\Big]^{\frac{1}{2}}\\
 =&\Big[\frac{1-f^\omega_s(\lambda,1-\lambda)-f^\omega_s(1-\lambda,\lambda)}{2}
 \Big]^{\frac{1}{2}},
\end{aligned}
\end{equation*}
where the minimum  is  achieved under the orthonormal basis $\{|0\rangle,|1\rangle\}$ of subsystem  $\mathcal{H}_{1}$.

By substituting Eqs. (\ref{example23}), (\ref{example231}) and (\ref{example232}) into Eq. (\ref{example211}), we can obtain
\begin{equation}\label{example241}
\begin{aligned}
 \mathcal{F}^2_f(\rho)=&\min\limits_{\{|\chi_l\rangle\}} \dfrac{f(0)}{4}\dfrac{(2\lambda-1)^2}{\lambda f(\frac{1-\lambda}{\lambda})}\sum\limits_{l=1}^{2}(2a_l-1)^2+\sum\limits_{l=1}^{2}a_l(1-a_l)\\
 =&\min\limits_{\{|\chi_l\rangle\}}2\Big(\dfrac{f(0)(2\lambda-1)^2}{\lambda f(\frac{1-\lambda}{\lambda})}-1\Big)(a_1-\frac{1}{2})^2+\frac{1}{2}.
\end{aligned}
\end{equation}
When $\dfrac{f(0)(2\lambda-1)^2}{\lambda f(\frac{1-\lambda}{\lambda})}>1$,  $\mathcal{F}_f(\rho)=\sqrt{\frac{1}{2}}$ where the minimum  is  achieved under the orthonormal basis $\{\frac{|0\rangle+|1\rangle}{\sqrt{2}},\frac{|0\rangle-|1\rangle}{\sqrt{2}}\}$ of subsystem  $\mathcal{H}_{1}$.
When $\dfrac{f(0)(2\lambda-1)^2}{\lambda f(\frac{1-\lambda}{\lambda})}<1$,  $\mathcal{F}_f(\rho)=\sqrt{\dfrac{f(0)(2\lambda-1)^2}{2\lambda f(\frac{1-\lambda}{\lambda})}}$ where the minimum  is  achieved under the orthonormal basis $\{|0\rangle,|1\rangle\}$ of subsystem  $\mathcal{H}_{1}$.

 For the indicator $\mathcal{I}^\chi_{s,\omega}(\rho)$ defined as (\ref{correlation2}), when the nondegenerate spectrum  $\chi$ of subsystem  $\mathcal{H}_{1}$ is $\chi=\{1,-1\}$, we can get
\begin{equation}\label{example26}
\begin{aligned}
 \mathcal{I}^\chi_{s,\omega}(\rho)=&1-\max\limits_{\{\Pi_\chi\}}\Big[\lambda|\langle\psi_1|\Pi_\chi\otimes\textbf{1}_2|\psi_1\rangle|^2
 +(1-\lambda)|\langle\psi_2|\Pi_\chi\otimes\textbf{1}_2|\psi_2\rangle|^2\\
 &+f^\omega_s(\lambda,1-\lambda)|\langle\psi_1|\Pi_\chi\otimes\textbf{1}_2|\psi_2\rangle|^2
 +f^\omega_s(1-\lambda,\lambda)|\langle\psi_2|\Pi_\chi\otimes\textbf{1}_2|\psi_1\rangle|^2\Big].
\end{aligned}
\end{equation}
Since the Hermitian operator $\Pi_\chi$  of subsystem  $\mathcal{H}_{1}$ has fixed nondegenerate spectrum  $\chi=\{1,-1\}$,
there exits a unitary operator $U$ such that $\Pi_\chi=U(|0\rangle\langle0|-|1\rangle\langle1|)U^\dagger=|\alpha_{\Pi_\chi}\rangle\langle\alpha_{\Pi_\chi}|-|\beta_{\Pi_\chi}\rangle\langle\beta_{\Pi_\chi}|$
with $|\alpha_{\Pi_\chi}\rangle=U|0\rangle$ and $|\beta_{\Pi_\chi}\rangle=U|1\rangle$.
After calculation, we have
\begin{equation}\label{example27}
\begin{aligned}
\langle\psi_1|\Pi_\chi\otimes\textbf{1}_2|\psi_1\rangle=\langle\psi_2|\Pi_\chi\otimes\textbf{1}_2|\psi_2\rangle=0,
\end{aligned}
\end{equation}
and
\begin{equation}\label{example28}
\begin{aligned}
\langle\psi_1|\Pi_\chi\otimes\textbf{1}_2|\psi_2\rangle
=\frac{|\langle\alpha_{\Pi_\chi}|0\rangle|^2
-|\langle\beta_{\Pi_\chi}|0\rangle|^2-|\langle\alpha_{\Pi_\chi}|1\rangle|^2+|\langle\beta_{\Pi_\chi}|1\rangle|^2}{2}
=a_{\Pi_\chi}-b_{\Pi_\chi}
\end{aligned}
\end{equation}
with $a_{\Pi_\chi}=|\langle\alpha_{\Pi_\chi}|0\rangle|^2$ and $b_{\Pi_\chi}=|\langle\beta_{\Pi_\chi}|0\rangle|^2$.
By Eqs. (\ref{example26}), (\ref{example27}) and (\ref{example28}), there is
\begin{equation*}
\begin{aligned}
\mathcal{I}^\chi_{s,\omega}(\rho)
=&1-[f^\omega_s(\lambda,1-\lambda)
 +f^\omega_s(1-\lambda,\lambda)]\max\limits_{\{\Pi_\chi\}}(a_{\Pi_\chi}-b_{\Pi_\chi})^2\\
 =&1-[f^\omega_s(\lambda,1-\lambda)+f^\omega_s(1-\lambda,\lambda)].
\end{aligned}
\end{equation*}
Here the second equality holds because
\begin{equation*}
\begin{aligned}
a_{\Pi_\chi}+b_{\Pi_\chi}=&|\langle\alpha_{\Pi_\chi}|0\rangle|^2+|\langle\beta_{\Pi_\chi}|0\rangle|^2
=1,
\end{aligned}
\end{equation*}
and $\max\limits_{\{\Pi_\chi\}}(2a_{\Pi_\chi}-1)^2=1$
with the maximum being  achieved when $a_{\Pi_\chi}=0$ and $b_{\Pi_\chi}=1$  (that is, $|\alpha_{\Pi_\chi}\rangle=|1\rangle$ and $|\beta_{\Pi_\chi}\rangle=|0\rangle$ ) or $a_{\Pi_\chi}=1$ and $b_{\Pi_\chi}=0$
 (that is, $|\alpha_{\Pi_\chi}\rangle=|0\rangle$ and $|\beta_{\Pi_\chi}\rangle=|1\rangle$ ).
So far, we have deduced that $\mathcal{I}^\chi_{s,\omega}(\rho)$ of quantum state $\rho$ defined as (\ref{example20}),
\begin{equation*}
\begin{aligned}
\mathcal{I}^\chi_{s,\omega}(\rho)=&\min\limits_{\{\Pi_\chi\}} I^\omega_s(\rho,\Pi_\chi\otimes \textbf{1}_2)
=1-f^\omega_s(\lambda,1-\lambda)-f^\omega_s(1-\lambda,\lambda),
\end{aligned}
\end{equation*}
where the minimum  is  achieved when  $|\alpha_{\Pi_\chi}\rangle=|1\rangle$ and $|\beta_{\Pi_\chi}\rangle=|0\rangle$ or
$|\alpha_{\Pi_\chi}\rangle=|0\rangle$ and $|\beta_{\Pi_\chi}\rangle=|1\rangle$, that is, $\Pi_\chi=|0\rangle\langle0|-|1\rangle\langle1|$
or $\Pi_\chi=|1\rangle\langle1|-|0\rangle\langle0|$.

\section{The relationship with entanglement measure }

\subsection{ The relationship between $\mathcal{I}^\omega_s(\rho)$ (or $\mathcal{F}_f(\rho)$ ) and entanglement measure }

For a bipartite pure state $\rho=|\psi\rangle\langle\psi|$ of the quantum system $\mathcal{H}_{1}\otimes \mathcal{H}_{2}$ with dim$\mathcal{H}_{i}=d_i$, if the Schmidt decomposition of pure state $|\psi\rangle$ is $|\psi\rangle=\sum\limits_i\lambda_i|i_1\rangle|i_2\rangle$, then $\mathcal{I}^\omega_s(|\psi\rangle\langle\psi|)=\mathcal{F}_f(|\psi\rangle\langle\psi|)=\sqrt{1-\sum\limits_i\lambda_i^4}$ by Example 1.
 In addition, we can also get $I$-concurrence \cite{RungtaBuzekCavesHilleryMilburn2001} $C(|\psi\rangle):=\sqrt{2[1-\textrm{Tr}(\rho_1^2)]}=\sqrt{2(1-\sum\limits_i\lambda_i^4)}$.
So far, for bipartite pure states, we can established the relationship between the indicator $ \mathcal{I}^\omega_s(\rho)$  (or $\mathcal{F}_f(\rho)$)  and  $I$-concurrence $C(|\psi\rangle)$ as follows:

$\emph{Proposition 1.}$  For any bipartite pure state $\rho=|\psi\rangle\langle\psi|$ of the quantum system $\mathcal{H}_{1}\otimes \mathcal{H}_{2}$,
 $ \mathcal{I}^\omega_s(\rho)$, $\mathcal{F}_f(\rho)$ and  $I$-concurrence $C(|\psi\rangle)$ are equivalent except for a constant $\sqrt{2}$, that is,
$\mathcal{I}^\omega_s(|\psi\rangle\langle\psi|)=\mathcal{F}_f(|\psi\rangle\langle\psi|)=\frac{C(|\psi\rangle)}{\sqrt{2}}$ where $I$-concurrence
$C(|\psi\rangle):=\sqrt{2[1-\textrm{Tr}(\rho_1^2)]}$.

Based on Proposition 1 and
 the fact that $I$-concurrence $C(|\psi\rangle)$ is  not increase on average under local operation and classical
communication (LOCC) \cite{HorodeckiRMP2009},  $ \mathcal{I}^\omega_s(|\psi\rangle\langle\psi|)$  and $\mathcal{F}_f(|\psi\rangle\langle\psi|)$ are   not increase on average under LOCC for bipartite pure states. Hence, we can obtain the following conclusion.

$\emph{Proposition 2.}$  $ \mathcal{I}^\omega_s(|\psi\rangle\langle\psi|)$ and $\mathcal{F}_f(|\psi\rangle\langle\psi|)$ are  not increase on average under LOCC for bipartite pure states, that is,
\begin{equation*}
\begin{aligned}
\mathcal{I}^\omega_s(|\psi\rangle\langle\psi|)\geq\sum\limits_ip_i\mathcal{I}^\omega_s(|\varphi_i\rangle\langle\varphi_i|),
\mathcal{F}_f(|\psi\rangle\langle\psi|)\geq\sum\limits_ip_i\mathcal{F}_f(|\varphi_i\rangle\langle\varphi_i|),
\end{aligned}
\end{equation*}
where $\{p_i,|\varphi_i\rangle\}$ is the output ensemble of the pure state $|\psi\rangle$ after  LOCC.

For any bipartite pure state $|\psi\rangle$,  it is a classical-quantum state if and only if it is a separable state.
For any bipartite pure state $|\psi\rangle$,
$\mathcal{I}^\omega_s(|\psi\rangle\langle\psi|)$  and $\mathcal{F}_f(|\psi\rangle\langle\psi|)$  satisfy some conditions of entanglement measure:
 $\mathcal{I}^\omega_s(|\psi\rangle\langle\psi|)=0$  (or $\mathcal{F}_f(|\psi\rangle\langle\psi|)=0$) if and only if the pure state $|\psi\rangle$ is a separable state by (P1) of Theorem 1; $\mathcal{I}^\omega_s(|\psi\rangle\langle\psi|)$ (or $\mathcal{F}_f(|\psi\rangle\langle\psi|)$) is invariant under local unitary transformation by (P2) of Theorem 1;
$\mathcal{I}^\omega_s(|\psi\rangle\langle\psi|)$ (or $\mathcal{F}_f(|\psi\rangle\langle\psi|)$) is  not increase on average under LOCC for bipartite pure states by  Proposition 2. Hence, we  conclude the following:

$\emph{Proposition 3.}$ The indicator $ \mathcal{I}^\omega_s(|\psi\rangle\langle\psi|)$  and $\mathcal{F}_f(|\psi\rangle\langle\psi|)$ reduce to  entanglement measure when $\rho=|\psi\rangle\langle\psi|$ is a bipartite pure state.

In a bipartite quantum system  $\mathcal{H}_{1}\otimes \mathcal{H}_{2}$ with dim$\mathcal{H}_{i}=d_i$, let $E_1(|\psi\rangle):=\sqrt{2}\mathcal{I}^\omega_s(|\psi\rangle\langle\psi|)$ when the quantum state is a pure state,
and $E_1(\rho):=\sqrt{2}\min\limits_{\{p_i,|\psi^{(i)}\rangle\}}\sum\limits_ip_i\mathcal{I}^\omega_s(|\psi^{(i)}\rangle\langle\psi^{(i)}|)$ when
the quantum state $\rho$ is a mixed state where the  minimum  is  taken over  all  ensemble
decompositions  $\{p_i,|\psi^{(i)}\rangle\}$ of $\rho$. By Proposition 1, $E_1(\rho)$ is exactly $I$-concurrence of the mixed state $\rho$
defined via convex roof construction \cite{MintertKusBuchleitner2004}.

\subsection{ The relationship between $\mathcal{I}^\chi_{\omega,s}(\rho)$  and entanglement measure }

From property (1) of the generalized Wigner-Yanase skew information,  $I^\omega_{s}(\rho,X)=V(\rho,X)$  when $\rho$ is the pure state. Hence, when $\rho$ is a pure state, all  $I^\omega_{s}(\rho,X)$ are equal for any $s\leq0,0<\omega<1$, namely,
when $\rho$ is a pure state, $I^\omega_{s}(\rho,X)=I^{\omega'}_{s'}(\rho,X)$ for any $s\leq0,s'\leq0,0<\omega<1,0<\omega'<1$.
Since $\mathcal{I}^\chi_{{\frac{1}{2}},0}(\rho)$ is an entanglement measure for bipartite pure state \cite{GirolamiTufarelliAdesso2013}, so we can obtain the indicator $\mathcal{I}^\chi_{\omega,s}(\rho)$  defined as  (\ref{correlation2}) is also an entanglement measure for bipartite pure state.

$\emph{Proposition 4.}$ The indicator $ \mathcal{I}^\chi_{\omega,s}(\rho)$  defined as  (\ref{correlation2})  reduces to an entanglement measure when $\rho=|\psi\rangle\langle\psi|$ is a bipartite pure state.

For a bipartite pure state $|\psi\rangle$ of the quantum system $\mathcal{H}_{1}\otimes \mathcal{H}_{2}$ with dim$\mathcal{H}_{1}=2$ and dim$\mathcal{H}_{2}=d$, when nondegenerate spectrum  $\chi$ of subsystems  $\mathcal{H}_{1}$ is $\chi=\{-1,1\}$, we have $\mathcal{I}^\chi_{\omega,s}(|\psi\rangle\langle\psi|)=2(1-\textrm{Tr}\rho_1^2)$  by Example 1. Then we can establish the relationship between indicator $ \mathcal{I}^\chi_{\omega,s}(|\psi\rangle\langle\psi|)$  defined as  (\ref{correlation2})  and  $I$-concurrence $C(|\psi\rangle)$ \cite{RungtaBuzekCavesHilleryMilburn2001}.

$\emph{Proposition 5.}$  For any  bipartite pure state $|\psi\rangle$ of the quantum system $\mathcal{H}_{1}\otimes \mathcal{H}_{2}$ with dim$\mathcal{H}_{1}=2$ and dim$\mathcal{H}_{2}=d$,
the indicator $ \mathcal{I}^\chi_{\omega,s}(|\psi\rangle\langle\psi|)$  defined as  (\ref{correlation2}) and  $I$-concurrence $C(|\psi\rangle)$ obey
$\sqrt{\mathcal{I}^\chi_{\omega,s}(|\psi\rangle\langle\psi|)}=C(|\psi\rangle)$ where nondegenerate spectrum is $\chi=\{-1,1\}$ and
$C(|\psi\rangle)=\sqrt{2[1-\textrm{Tr}(\rho_1^2)]}$.

In a bipartite quantum system  $\mathcal{H}_{1}\otimes \mathcal{H}_{2}$ with dim$\mathcal{H}_{1}=2$ and dim$\mathcal{H}_{2}=d$, let $E_2(|\psi\rangle):=\sqrt{\mathcal{I}^\chi_{\omega,s}(|\psi\rangle\langle\psi|)}$ with nondegenerate spectrum being $\chi=\{-1,1\}$ when the quantum state is a pure state,
and $E_2(\rho):=\min\limits_{\{p_i,|\psi^{(i)}\rangle\}}\sum\limits_i p_i\sqrt{\mathcal{I}^\chi_{\omega,s}(|\psi^{(i)}\rangle\langle\psi^{(i)}|)}$ when
the quantum state $\rho$ is a mixed state with the  minimum    being taken over all possible  ensemble
decompositions  $\{p_i,|\psi^{(i)}\rangle\}$ of $\rho$. By Proposition 5, $E_2(\rho)$ is  $I$-concurrence of the mixed state $\rho$
defined via convex roof construction \cite{MintertKusBuchleitner2004}.

\section{Discussion}
 In this paper, the  generalized Wigner-Yanase skew information we have studied, which possesses some desirable characteristics, serves as an essential tool for quantifying bipartite nonclassical correlation. Based on two different local observables, we  first propose the  indicators $\mathcal{I}^\omega_s(\rho)$, $\mathcal{F}_f(\rho)$ and $\mathcal{I}^\chi_{\omega,s}(\rho)$ to quality nonclassical correlations.  These indicators have some good properties.
This means that we are able to quantify nonclassical correlation, and we have verified this through specific examples. For bipartite pure states, we not only find that both of these indicators reduce to entanglement measure, but also discover that they are both related to the entanglement measure $I$-concurrence (that is, $\sqrt{2}\mathcal{I}^\omega_s(|\psi\rangle\langle\psi|)=\sqrt{2}\mathcal{F}^\omega_f(|\psi\rangle\langle\psi|)=C(|\psi\rangle)$ for any bipartite pure states, $\sqrt{\mathcal{I}^\chi_{\omega,s}(|\psi\rangle\langle\psi|)}=C(|\psi\rangle)$ for the bipartite pure states of the quantum system $\mathcal{H}_{1}\otimes \mathcal{H}_{2}$ with dim$\mathcal{H}_{1}=2$ and dim$\mathcal{H}_{2}=d$ and  $\chi=\{-1,1\}$).
Using the indicator $\mathcal{I}^\omega_s(|\psi\rangle\langle\psi|)$ in pure states, we  construct $I$-concurrence for arbitrary bipartite mixed states through the method of convex roof construction, and a similar approach can be applied to the  indicator for the bipartite mixed states of the quantum system $\mathcal{H}_{1}\otimes \mathcal{H}_{2}$ with dim$\mathcal{H}_{1}=2$, dim$\mathcal{H}_{2}=d$ and  $\chi=\{-1,1\}$).
This further reveals the close connection between these two indicators and entanglement measure.

There are some open questions for this paper. First, we  only find the decreasing under CPTP for $\mathcal{I}^\omega_s(\rho)$ and $\mathcal{I}^\chi_{\omega,s}(\rho)$ when $s=0$ or when $\omega=\frac{1}{2}$ and $ s=-1$, and further investigation is needed for other scenarios. Second, we have only discussed the relationship between the second indicator $\mathcal{I}^\chi_{\omega,s}(\rho)$  and $I$-concurrence when $\rho\in\mathcal{H}_{1}\otimes \mathcal{H}_{2}$ with dim$\mathcal{H}_{1}=2$, dim$\mathcal{H}_{2}=d$ and  $\chi=\{-1,1\}$, and we need to continue exploring this relationship for bipartite quantum states of arbitrary dimensions.

{\bf Funding}
This work was supported by  the National Natural Science Foundation of China under Grant No. 11701135, the Hebei
Natural Science Foundation of China under Grant No. A2017403025, supported by National Pre-research Funds of Hebei GEO University (Grant KY202316 and KY2025YB15), PhD Research Startup Foundation of Hebei GEO University (Grant BQ201615).

{\bf Data Availability Statement}
No data was used for the research described in the article.

{\bf Declaration of competing interest}
The authors declare that they have no known competing financial interests or personal relationships that could have appeared to influence the work reported in this paper.

\appendix
\section{The proof of properties for the generalized Wigner-Yanase skew information }

(1)  The generalized Wigner-Yanase skew information $I^\omega_{s}(\rho,X)$ is  monotonically decreasing with  $s$ because the function $f^\omega_s(a,b)$ is monotonically increasing with respect to $s$ \cite{Kuang2021}.
When the quantum state is pure state, $I^\omega_{s}(|\psi\rangle\langle\psi|,X)= V(|\psi\rangle\langle\psi|,X)=\langle\psi|X^2|\psi\rangle-\langle\psi|X|\psi\rangle^2=F_f(|\psi\rangle\langle\psi|,X)$ for any $s$ and $0<\omega<1$.

(2) Let the spectral decomposition of quantum state $\rho$ be $\rho=\sum\limits_i\lambda_i|\psi_i\rangle\langle\psi_i|$.
According to the definition of the generalized Wigner-Yanase skew information, we can get
\begin{equation}\label{fP1}
\begin{aligned}
I^\omega_s(\rho,X)=&\sum\limits_{i\neq j}[\lambda_i-f^\omega_s(\lambda_i,\lambda_{j})]|\langle\psi_i|X|\psi_{j}\rangle|^2\\
=&\sum\limits_{i< j}[\lambda_i+\lambda_{j}-f^\omega_s(\lambda_i,\lambda_{j})-f^\omega_s(\lambda_{j},\lambda_{i})]|\langle\psi_i|X|\psi_{j}\rangle|^2\\
=&\sum\limits_{i< j}[\omega\lambda_i+(1-\omega)\lambda_{j}+\omega\lambda_{j}+(1-\omega)\lambda_i-f^\omega_s(\lambda_i,\lambda_{j})
-f^\omega_s(\lambda_{j},\lambda_{i})]|\langle\psi_i|X|\psi_{j}\rangle|^2.
\end{aligned}
\end{equation}
 When $s\leq0$, we can  derive as follows:
\begin{align}
\omega\lambda_i+(1-\omega)\lambda_{j}-f^\omega_s(\lambda_i,\lambda_{j})
\geq &\omega\lambda_i+(1-\omega)\lambda_{j}-f^\omega_0(\lambda_i,\lambda_{j})\label{fp2}\\
= &\omega\lambda_i+(1-\omega)\lambda_{j}-\lambda_i^\omega\lambda_{j}^{(1-\omega)}
\geq 0\label{fp3},
\end{align}
if $\lambda_i\lambda_j\neq0$;
and
\begin{align}
\omega\lambda_i+(1-\omega)\lambda_{j}-f^\omega_s(\lambda_i,\lambda_{j})
= &\omega\lambda_i+(1-\omega)\lambda_{j}-0
\geq 0\label{}\nonumber,
\end{align}
if $\lambda_i\lambda_j=0$.
Here  inequality (\ref{fp2}) holds because  the function $f^\omega_s(a,b)$ is monotonically increasing with respect to $s$ \cite{Kuang2021},
and inequality (\ref{fp3}) follows from  inequality $a^xb^{1-x}\leq xa+(1-x)b$ with $a>0,b>0,0< x<1$ \cite{Kuang2021}.
In conclusion, when $s\leq0$, we can get
\begin{equation}\label{fP5}
\begin{aligned}
\omega\lambda_i+(1-\omega)\lambda_{j}-f^\omega_s(\lambda_i,\lambda_{j})\geq 0.
\end{aligned}
\end{equation}
Similarly, we can also obtain
\begin{equation}\label{fP6}
\begin{aligned}
\omega\lambda_{j}+(1-\omega)\lambda_{i}-f^\omega_s(\lambda_{j},\lambda_{i})\geq 0.
\end{aligned}
\end{equation}
Using inequalities (\ref{fP1}), (\ref{fP5}), and (\ref{fP6}), we have $I^\omega_s(\rho,X)\geq0$ when $s\leq0$.

 If the quantum state $\rho$ and the observable $X$ are commutative, then the observable $X$ can be diagonalized in the eigenspace of the quantum state $\rho$. Hence, $I^\omega_s(\rho,X)=\sum\limits_{i\neq j}[\lambda_i-f^\omega_s(\lambda_i,\lambda_{j})]|\langle\psi_i|X|\psi_{j}\rangle|^2=0$.
If $I^\omega_s(\rho,X)=0$, then $I^\omega_0(\rho,X)=0$ because the generalized Wigner-Yanase skew information $I^\omega_{s}(\rho,X)$ is  monotonically decreasing with  $s$ and  $I^\omega_0(\rho,X)\geq0$. The Wigner-Yanase-Dyson skew information takes a value of zero if and only if $\rho$ and  $X$ are commutative.
Since the generalized Wigner-Yanase skew information $I^\omega_0(\rho,X)$ is Wigner-Yanase-Dyson skew information when $s=0$, so we can get $\rho$ and  $X$ are commutative.

(3) Assuming that the spectral decomposition of quantum state $\rho$ is $\rho=\sum\limits_i\lambda_i|\psi_i\rangle\langle\psi_i|$ with $\lambda_i$ being the eigenvalue and $|\psi_i\rangle$ being orthonormal bases, then the spectral decomposition of quantum state $U\rho U^\dagger$ is $U\rho U^\dagger=\sum\limits_i\lambda_iU|\psi_i\rangle\langle\psi_i|U^\dagger$ where $\lambda_i$ is the eigenvalue and $U|\psi_i\rangle$ is orthonormal bases. Using the definition of the generalized Wigner-Yanase skew information, we can derive that the generalized Wigner-Yanase skew information $I^\omega_s(\rho ,X)$ is unitary invariant, that is,
\begin{equation*}
\begin{aligned}
I^\omega_s(U\rho U^\dagger,X)=\sum\limits_{i\neq j}[\lambda_i-f^\omega_s(\lambda_i,\lambda_{j})]|\langle\psi_i|U^\dagger X U|\psi_{j}\rangle|^2
=I^\omega_s(\rho, U^\dagger X U ).
\end{aligned}
\end{equation*}

(4) Let the spectral decomposition of quantum state $\rho$ be $\rho=\sum\limits_i\lambda_i|\psi_i\rangle\langle\psi_i|$.
 From inequalities (\ref{fP1}), we can find that
\begin{equation*}
\begin{aligned}
I^\omega_{-1}(\rho,X)=\sum\limits_{i< j}[\lambda_i+\lambda_{j}-f^\omega_{-1}(\lambda_i,\lambda_{j})
-f^\omega_{-1}(\lambda_{j},\lambda_{i})]|\langle\psi_i|X|\psi_{j}\rangle|^2.
\end{aligned}
\end{equation*}
When $\lambda_i\lambda_{j}\neq0$, we can obtain that
\begin{equation*}
\begin{aligned}
-f^\omega_{-1}(\lambda_i,\lambda_{j})-f^\omega_{-1}(\lambda_{j},\lambda_{i})
=\dfrac{\lambda_i\lambda_{j}(\lambda_i+\lambda_{j})}{(\lambda_i-\lambda_{j})^2(\omega^2-\omega)-\lambda_i\lambda_{j}}.
\end{aligned}
\end{equation*}
When $0<\omega\leq\frac{1}{2}$,  $-f^\omega_{-1}(\lambda_i,\lambda_{j})-f^\omega_{-1}(\lambda_{j},\lambda_{i})$ is monotonically increasing with  $\omega$; when $\frac{1}{2}\leq\omega<1$,  $-f^\omega_{-1}(\lambda_i,\lambda_{j})-f^\omega_{-1}(\lambda_{j},\lambda_{i})$ is monotonically decreasing with  $\omega$. Hence, when $0<\omega\leq\frac{1}{2}$, $I^\omega_{-1}(\rho,X)$ is monotonically increasing with  $\omega$;
when $\frac{1}{2}\leq\omega<1$,  $I^\omega_{-1}(\rho,X)$ is monotonically decreasing with  $\omega$. So,
\begin{equation}\label{fP11}
\begin{aligned}
I^\omega_{-1}(\rho,X)\leq I^{\frac{1}{2}}_{-1}(\rho,X).
\end{aligned}
\end{equation}
For $\rho=\sum\limits_jq_j|\phi_j\rangle\langle\phi_j|$, when $-1\leq s\leq0$, we have
\begin{equation*}
\begin{aligned}
I^\omega_s(\rho,X)\leq I^\omega_{-1}(\rho,X)\leq I^{\frac{1}{2}}_{-1}(\rho,X)\leq\sum\limits_jq_jI^{\frac{1}{2}}_{-1}(|\phi_j\rangle\langle\phi_j|,X)
=\sum\limits_jq_jI^\omega_s(|\phi_j\rangle\langle\phi_j|,X),
\end{aligned}
\end{equation*}
where the first inequality holds because  the generalized Wigner-Yanase skew information $I^\omega_{s}(\rho,X)$ is  monotonically decreasing with  $s$,  the second inequality because of inequalities (\ref{fP11}), the third
inequality follows from the facts $I^{\frac{1}{2}}_{-1}(\rho,X)$ reduces to quantum Fisher information when $\omega=\frac{1}{2}$ and $s=-1$, and quantum Fisher information is convex. When the quantum state is pure state, $I^\omega_{s}(|\psi\rangle\langle\psi|,X)= V(|\psi\rangle\langle\psi|,X)=\langle\psi|X^2|\psi\rangle-\langle\psi|X|\psi\rangle^2$ for any $s$ and $0<\omega<1$, so the first equality is true.

(5) If function $m_s(\lambda_i,\lambda_{j})$ of Ref. \cite{YangQiao2022} is replaced by the  $f^\omega_s(\lambda_i,\lambda_{j})$ and the facts (that is, $f^\omega_s(ta,tb)=tf^\omega_s(a,b)$ and $f^\omega_s(a,a)=a$) are utilized, we can demonstrate additivity of
the generalized Wigner-Yanase skew information by a similar approach of Ref. \cite{YangQiao2022}.

\end{document}